\documentstyle[preprint,aps]{revtex}
\tighten
\begin{document}
\draft
\title{Hadronic $B$ decays: Supersymmetric enhancement and 
a simple spectator model}
\author{W. N. Cottingham and H. Mehrban }
\address{H H Wills Physics Laboratory,\\ 
Royal Fort, Tyndall Ave, Bristol, BS8 1TL, UK}
\author{I. B. Whittingham}
\address{School of Computer Science, Mathematics and Physics,\\ 
James Cook University, Townsville, Australia, 4811}
\date{\today}
\maketitle
\title{Abstract}
\begin{abstract}
Two aspects of hadronic $B$ decays are investigated. Firstly, 
the supersymmetric enhancement of hadronic $b$ decays $b \rightarrow qg 
\rightarrow q q^{\prime} \bar{q}^{\prime},\;q \in \{d,s\}$ by gluino
penguin processes is studied through their effect on the Wilson 
coefficients of the effective Hamiltonian. The gluino penguin is dominated
by the magnetic dipole transition which is strongly magnified relative 
to the electric monopole driven standard model gluon penguin by the 
renormalization-induced QCD corrections, resulting in quark decay rates
for pure penguin processes which, at scales $O(m_{b})$, can greatly exceed
the standard model rates. The $CP$ asymmetries are, however, relatively
unaffected. Secondly, hadronization of the final state quarks is studied
through a simple phase space spectator model. We consider two extreme 
models for color flow during meson formation; one in which color flow
is ignored and one of color suppression in which low mass 
meson formation occurs
only for color singlet quark antiquark pairs.  We find that processes in
which the spectator antiquark $\bar{q}_{s}$ combines with $q^{\prime}$ are
relatively insensitive to the color flow model whereas processes in which
$\bar{q}_{s}$ combines with $q$ are very sensitive to color suppression.
\end{abstract}
\narrowtext
\section{Introduction}
These are exciting times for $B$ physics. Recently the CLEO collaboration
\cite{CLEO} reported the first measurements of a number of exclusive
charmless hadronic $B$ decays which provide conclusive evidence for
hadronic penguin processes, and the next decade will see \cite{Babar}
intensive investigation of the $B$-meson system at CESR, the Tevatron, 
HERA, the SLAC and KEK $B$-factories, and at the LHC.
In particular, measurements of rare flavor-changing $B$ decays \cite{Buras98} 
will provide windows for
the discovery of indirect effects of new physics beyond the standard
model (SM). For example, data on CP asymmetries should indicate whether 
the Cabbibo-Kobayashi-Maskawa (CKM) CP violation mechanism of the SM 
is correct or whether new sources of CP violation are needed.  The most
favoured candidate theory for this new physics is supersymmetry (SUSY) 
for which rare $B$ decays offer possible insight because 
\cite{Gronau97,Hewett97} the measured observables involve SM and SUSY 
processes occurring at the same order of perturbation theory. 
In particular, the $b \rightarrow s$ transition provides an opportunity 
to study CP violation from non-standard phases \cite{Atwood97}. 
Probing for SUSY in $b \rightarrow s \gamma $ decays has been examined
by many authors 
\cite{Hewett97,Bert91,Kagan98,Neubert98,Ciuchini98,Chua98,Aoki98} 
and there is also significant current interest in the 
$b \rightarrow sg$ penguin decay \cite{Hou97a,Hou99,Desh98,He98} 
for which it has 
been argued \cite{Hou97} that enhancement for on-shell gluons is needed
from non-SM physics to explain the CLEO measurement \cite{CLEO} of a large
branching ratio for the inclusive process 
$B \rightarrow \eta^{\prime}X_{s}$ and the 
$\eta^{\prime}-g-g$ gluon anomaly.

The most predictive of the SUSY models is the (constrained) minimally 
supersymmetric standard model (MSSM) \cite{MSSM,Abel96} based on 
spontaneously broken $N=1$ supergravity with flat K\"{a}hler metrics 
\cite{SUGRA}, universal explicit soft-SUSY breaking terms at the scale
$M_{\text{MSSM}}\sim M_{U}$ and spontaneous breaking of the 
$SU(2)\otimes U(1)$ symmetry driven by radiative corrections. The 
unification scale $M_{U}$ boundary conditions 
are a universal scalar mass $m_{0}$,
universal gaugino mass $m_{1/2}$ and a universal soft SUSY-breaking
trilinear scalar coupling $A$. After minimization of the full 
one-loop Higgs effective potential the MSSM is represented by a  
four-dimensional parameter space $\{m_{0},m_{1/2},A,\tan \beta \}$,
where $\tan \beta \equiv v_{2}/v_{1}$ is the ratio of the vacuum
expectation values of the two Higgs fields, together with the sign of the 
coupling $\mu $ between the two Higgs fields.

In an earlier paper \cite{ACW98} we examined the SM and MSSM 
predictions at the weak scale quark level for the penguin
mediated decays $b \rightarrow qq^{\prime}\bar{q}^{\prime},q\in \{d,s\}$. 
For the SM we presented the relative contributions of the 
internal $u$, $c$ and $t$ quarks 
to the  gluon penguin and stressed the invalidity of the widespread 
assumption that the process is dominated by the $t$ quark 
\cite{Fleischer}. We also considered the relative magnitudes of the various
form factors and the role of the strong and weak phases 
\cite{Band79,Gerard91} and found, for example, that the CP violating 
phases for the $b \rightarrow 
d g$ and $b \rightarrow s g $ electric form factors, which dominate the
decay amplitude, have no simple relationship with any angle of the 
unitarity triangle. For the MSSM we explored the allowed regions of the 
parameter space to locate those regions which gave the largest 
modifications to the SM results and found, in contrast to the SM, 
that the magnetic amplitude dominates the electric amplitude and that 
there
are large regions of the MSSM parameter space for which the magnetic 
amplitude is greater than that of the SM. For regions of high $\tan \beta $
(=48) and low $(m_{0},m_{1/2})$ ranging from (225,150) to (275,225) the SUSY 
enhancement of the gluon-mediated exclusive hadronic $b$ decays
can be at the several percent level and the SUSY penguin processes 
dominate the SUSY box processes.  Similar effects of large $\tan \beta $
enhancement occur for the photon magnetic dipole operator in 
$b \rightarrow s \gamma $ and there is significant current interest in its
implications for $B \rightarrow X_{s}\gamma$ and $B \rightarrow X_{s}\;
l^{+}l^{-}\;(l=e,\mu , \tau )$ \cite{Hewett97,Chua98,Aoki98}.

Calculations of weak decays of hadrons within the framework of the
effective Hamiltonian involve \cite{Gronau97,Buras99}: 
(i) the computation
of the quark level decays $b \rightarrow qq^{\prime}\bar{q}^{\prime}$ at
the electroweak scale $M_{W}$,
(ii) determination of the Wilson coefficients $C_{k}(M_{W})$ through
matching of this full theory onto the five-quark low energy effective
theory in which the $W^{\pm}$, $t$ and all SUSY particles heavier than
$M_{W}$ are integrated out,
(iii) evolution of the Wilson coefficients down to the low energy scale
$\mu \sim O(m_{b}) $ using renormalization group improved perturbation
theory \cite{Buch96}, thereby incorporating the important short distance QCD
corrections, and (iv) the calculation of hadronic matrix 
elements for the hadronization of the final-state quarks into particular
final states, typically evaluated using the factorization assumption
\cite{fact}. 

In this earlier paper \cite{ACW98}, the QCD corrections arising from 
renormalization of the short distance
results down from the electroweak scale to the scale $m_{b}$ were not included. 
It was argued that these effects were not 
likely to alter the finding that the magnetic
amplitude is dominant in the MSSM as the QCD induced mixing effects
\cite{Buch96} produce an enhancement of the magnetic dipole operators
in the $\Delta B =1$ effective Hamiltonian relative to the current-current
penguin operators associated with the electric monopole amplitude.
Furthermore, as noted by G\'{e}rard and Hou \cite{Gerard91},
use of the $q^{2}$-dependent SM form factor $F^{L}_{1}(q^{2})$ already 
incorporates the
dominant part of the QCD corrections for the current-current penguin
operators and, therefore, that the main effects of QCD corrections will be
the renormalization of the strong coupling constant from $\alpha_{s}(M_{W})$
to $\alpha_{s}(m_{b})$. This would have the effect of increasing 
the penguin decay rates of the SM by the factor $\eta ^{2}$, where
$\eta \equiv \alpha_{s}(m_{b})/\alpha_{s}(M_{W})\approx 1.82$, and also 
increasing the MSSM penguin amplitudes relative to those of the SM. 

In the present paper we address each of the stages of the 
calculation of $B$ decay rates. In section II we incorporate the MSSM 
penguin processes into the effective Hamiltonian  
and fully implement the renormalization-induced
QCD corrections. We confirm that the MSSM processes are enhanced relative
to those of the SM by these QCD corrections and that the dominant MSSM
processes are indeed the magnetic dipole interactions. In section III we
compare the quark-level decay rates calculated from the QCD corrected 
effective Hamiltonian with $\eta ^{2}\Gamma^{(0)}$, where $\Gamma^{(0)}$
are our weak scale decay rates \cite{ACW98} calculated using $q^{2}$-
dependent form factors, and find that the G\'{e}rard and Hou 
conjecture is generally valid to within 10\% for the SM but greatly
underestimates the decay rates for the SM+MSSM. The hadronization of the
quark level processes is studied in section IV where, as an alternative to
the widely used factorization models, we adopt a simple spectator model.
The results obtained from this spectator model are given in section V.
Section VI presents our discussions and conclusions.
The input numerical values used in our calculations for the CKM parameters 
and quark masses are given in an appendix.

\section{Gluino Penguins and the Effective Hamiltonian}
With the inclusion of gluino-mediated penguin processes from SUSY, 
the total QCD penguin amplitude to lowest order in $\alpha_{s}$ 
for the 
decay process $b \rightarrow q\;g 
\rightarrow q\;q^{\prime}\bar{q}^{\prime}$ is
\begin{equation}
\label{bqcd1}
M^{\text{Peng}}=-\frac{ig^{2}_{s}}{4 \pi^{2}}\;
[\bar{u}_{q}(p_{q})T^{a}\hat{\gamma}_{\mu}u_{b}(p_{b})]
[\bar{u}_{q^{\prime}}(p_{q^{\prime}})
\gamma^{\mu}T^{a}v_{\bar{q}^{\prime}}(p_{\bar{q}^{\prime}})]
\end{equation}
where
\begin{equation}
\label{bqcd2}
\hat{\gamma}_{\mu} \equiv \gamma_{\mu}[F^{L}_{1}(q^{2})P_{L}+
F^{R}_{1}(q^{2})P_{R}] 
+\frac{i\sigma_{\mu\nu}q^{\nu}}{q^{2}}[F^{L}_{2}
(q^{2})P_{L}+F^{R}_{2}(q^{2})P_{R}].
\end{equation}
Here $F_{1}$ and $F_{2}$ are the electric (monopole) and magnetic (dipole)
form factors, 
$q=p_{b}-p_{q}$ is the gluon momentum, $P_{L(R)}\equiv (1\mp \gamma_{5})/2$
 are the chirality
projection operators and $T^{a}\;(a=1,\ldots,8)$ are the $SU(3)_{c}$ 
generators normalized to $Tr(T^{a}T^{b})=\frac{1}{2}\delta^{ab}$.
In writing (\ref{bqcd1}) we have neglected the smaller MSSM 
$b \rightarrow q g$ penguin processes mediated \cite{Bert91} 
by charged Higgs boson, 
chargino and neutralino exchanges. The gluino penguin amplitude is
enhanced relative to that for these processes by both the factor 
$\alpha_{s}/\alpha_{W}$ and the additional $\tilde{g}-\tilde{g}-g$
coupling with its large color factor.

For the SM, $q^{2}$ is not small compared to $m_{u}^{2}$ and $m_{c}^{2}$
and we must retain the $q^{2}$ dependence of the $u$ and $c$ contributions
to $F^{L}_{1}(q^{2})$. In the limit $m_{q}=0$ we then have \cite{ACW98}
\begin{eqnarray}
\label{bqcd3}
F^{L}_{1}(q^{2})&=&\frac{G_{F}}{\sqrt{2}}[\sum_{i=u,c}\;V^{*}_{iq}V_{ib}
\;f_{1}(x_{i},q^{2})+V^{*}_{tq}V_{tb}\;f_{1}(x_{t})], \nonumber \\
F^{R}_{1}(0)&=&0,  \\
\label{bqcd4}
F^{L}_{2}(0)&=&0,\nonumber  \\ 
F^{R}_{2}(0)&=&\frac{G_{F}}{\sqrt{2}}m_{b}\sum_{i=u,c,t}\;
V^{*}_{iq}V_{ib}\;f_{2}(x_{i})
\end{eqnarray}
where $V$ is the CKM matrix and $x_{i}^{2}\equiv m_{i}^{2}/M_{W}^{2}$. 
Explicit expressions for $f_{1}(x_{i},q^{2})$ are given in \cite{ACW98} 
but are not needed in the present study (see later). The $q^{2}=0$
functions are \cite{Inami81,Hou88}
\begin{eqnarray}
\label{bqcd5}
f_{1}(x)&=&\frac{1}{12(1-x)^{4}}[18x-29x^{2}+10x^{3}+x^{4} 
 - (8-32x+18x^{2}) \ln x],  \\
f_{2}(x)&=&\frac{-x}{4(1-x)^{4}}[2+3x-6x^{2}+x^{3}+6x \ln x].
\end{eqnarray}
For small $x_{i}$, $f_{2}(x_{i})\approx -\frac{1}{2}x_{i}$ whereas 
$f_{1}(x_{i}) \approx -\frac{2}{3} \ln x_{i} $.

For the  MSSM form factors we have $q^{2} \ll m^{2}_{\tilde{d}_{j}}$,
where $m_{\tilde{d}_{j}}$ is the mass of the loop $\tilde{d}_{j}$
squark, and we can use the $q^{2}=0$ form factors \cite{ACW98,Gerard84,Bert87}
\begin{eqnarray}
\label{bqcd10}
F^{L}_{1}(0)&=&\sum_{j}\;\Lambda^{bq}_{LLj}[C_{2}(G)A(\tilde{x}_{j})
                +C_{2}(R)B(\tilde{x}_{j})],   \\
\label{bqcd11}
F^{R}_{1}(0)&=&\sum_{j}\;\Lambda^{bq}_{RRj}[C_{2}(G)A(\tilde{x}_{j})
                +C_{2}(R)B(\tilde{x}_{j})],  \\
\label{bqcd12}
F^{L}_{2}(0)&=&\sum_{j}\;\{[m_{q}\;\Lambda^{bq}_{LLj}
+m_{b}\;\Lambda^{bq}_{RRj}]  \nonumber  \\
      && \times [C_{2}(G)C(\tilde{x}_{j})-C_{2}(R)D(\tilde{x}_{j})] \nonumber  \\
      && +m_{\tilde{g}}\;\Lambda^{bq}_{LRj}[C_{2}(G)E(\tilde{x}_{j})
                -4C_{2}(R)C(\tilde{x}_{j})]\},   \\
\label{bqcd13}
F^{R}_{2}(0)&=&\sum_{j}\{[m_{b}\;\Lambda^{bq}_{LLj}+
   m_{q}\;\Lambda^{bq}_{RRj}]   \nonumber  \\
    &&  \times [C_{2}(G)C(\tilde{x}_{j})-C_{2}(R)D(\tilde{x}_{j})] \nonumber \\
    && +m_{\tilde{g}}\;\Lambda^{bq}_{RLj}[C_{2}(G)E(\tilde{x}_{j})
                -4C_{2}(R)C(\tilde{x}_{j})]\}.
\end{eqnarray}
where $m_{\tilde{g}}$ is the gluino mass and the coefficient
\begin{equation}
\label{bqcd14}
 \Lambda^{bq}_{ABj} \equiv - \frac{g_{s}^{2}}{4m_{\tilde{g}}^{2}}
V^{jq\;*}_{\tilde{d}B}\;V^{jb}_{\tilde{d}A}
\end{equation}
describes the rotation from the down-diagonal 
interaction states to the $\tilde{d}$ mass eigenstates at the 
$d-\tilde{d}-\tilde{g}$ vertices. Here $(A,B)$ are chirality indices, 
$C_{2}(G)=3$ and $C_{2}(R)=
\sum_{a}T^{a}T^{a}=4/3$ are $SU(3)$ Casimir invariants, $j=1,\ldots,6$ 
labels the $d$ squark mass eigenstates and $\tilde{x}_{j}\equiv
m^{2}_{\tilde{d}_{L_{j}}}/m^{2}_{\tilde{g}}$. 
The matrices $V_{\tilde{d}L}$ and
$V_{\tilde{d}R}$ are obtained from the $(6\times 6)$ matrix $V_{\tilde{d}}=
(V_{\tilde{d}L},V_{\tilde{d}R})^{T}$ which diagonalizes the 
$\tilde{d}$ mass$^{2}$ matrix. The MSSM functions are
\begin{eqnarray}
\label{bqcd15}
A(x)&=&\frac{1}{6(1-x)^{4}}[3-9x+9x^{2}-3x^{3}  
         +(1-3x^{2}+2x^{3})\ln x],  \\
\label{bqcd16}
B(x)&=&\frac{-1}{18(1-x)^{4}}[11-18x+9x^{2}-2x^{3}+6 \ln x],  \\
\label{bqcd17}
C(x)&=&\frac{-1}{4(1-x)^{3}}[1-x^{2} +2x \ln x],  \\
\label{bqcd18}
D(x)&=&\frac{-1}{6(1-x)^{4}}[2+3x-6x^{2}+x^{3}+6x \ln x],  \\
\label{bqcd19}
E(x) &= &\frac{-1}{(1-x)^{2}}[1-x+x\ln x] .
\end{eqnarray}

The SUSY masses and diagonalizing matrices needed to evaluate the MSSM
form factors at $M_{W}$ were obtained (for details see \cite{Abel96})
using two-loop MSSM renormalization group equations (RGEs) for the 
gauge and Yukawa couplings and one-loop MSSM RGEs for the other SUSY
parameters.
The magnitudes of the MSSM form factors satisfy 
$|F^{R}_{2}| > |F^{L}_{1}| \gtrsim |F^{L}_{2}| \gg |F^{R}_{1}|$ 
for all regions of the allowed parameter space apart from the narrow 
region $\tan \beta =2,\;m_{1/2}=150$ and $m_{0} \gtrsim 1000 $ where
$|F^{L}_{1}|$ is slightly smaller than $|F^{R}_{2}|$. 
Outside this region the ratio $|F^{R}_{2}|/|F^{L}_{1}| $ exceeds unity and
increases strongly with $\tan \beta $.  
The result that $F^{R}_{2}$ is the largest MSSM form factor indicates that, 
in contrast to the SM, the magnetic dipole transition dominates the $b$ decay
process in the MSSM. To compare with the SM, we note that the ratio of 
the largest MSSM and SM form factors is
$|F^{R}_{2}(\text{MSSM})|/|F^{L}_{1}(\text{SM})(q^{2}=0)| \leq 0.4$ GeV.

The amplitude (\ref{bqcd1}) can be written
\begin{equation}
\label{bqcd19a}
M^{\text{Peng}}=M^{\text{SM}}+M^{\text{MSSM}}
\end{equation}
where
\begin{eqnarray}
\label{bqcd20}
M^{\text{SM}}&=&-i\frac{G_{F}}{\sqrt{2}}\;\frac{\alpha_{s}(M_{W})}{\pi}
\{\frac{1}{8}[\sum_{i=u,c}V^{*}_{iq}V_{ib}f_{1}(x_{i},q^{2}) 
 +V^{*}_{tq}V_{tb}f_{1}(x_{t})]O_{P} \nonumber  \\ 
&& +\frac{m_{b}}{2}\sum_{i=u,c,t}V^{*}_{iq}V_{ib}f_{2}(x_{i})O_{8} \}
\end{eqnarray}
and
\begin{equation}
\label{bqcd19b}
M^{\text{MSSM}}=-i\frac{\alpha_{s}(M_{W})}{\pi}\{\frac{1}{8}
[F^{L}_{1}(0)O_{P}+F^{R}_{1}(0)\tilde{O}_{P}]   
+\frac{1}{2}[F^{L}_{2}(0)O_{8}+F^{R}_{2}(0)\tilde{O}_{8}]\}.
\end{equation}
Here
\begin{equation}
\label{bqcd19c}
O_{P}=O_{4}+O_{6}-\frac{1}{3}(O_{3}+O_{5})
\end{equation}
is a combination of the standard QCD penguin four-fermion operators
\cite{Buch96}
\begin{eqnarray}
O_{3,5}& \equiv & (\bar{q}_{\alpha}\gamma^{\mu}b_{\alpha})_{\text{V-A}}\;
\sum_{q^{\prime}}(\bar{q}^{\prime}_{\beta}\gamma_{\mu}
q^{\prime}_{\beta})_{\text{V} \mp \text{A}}  \\
O_{4,6}& \equiv & (\bar{q}_{\alpha}\gamma^{\mu}b_{\beta})_{\text{V-A}}\;
\sum_{q^{\prime}}(\bar{q}^{\prime}_{\beta}\gamma_{\mu}
q^{\prime}_{\alpha})_{\text{V} \mp \text{A}},  
\end{eqnarray}
where $q^{\prime} \in \{u,d,s,c,b\}$, and $O_{8}$ is 
the chromomagnetic dipole operator
\begin{equation}
\label{bqcd21}
O_{8}\equiv \frac{g_{s}^{2}}{8 \pi ^{2}}m_{b}[\bar{q}_{\alpha}
\sigma^{\mu \nu}(1+\gamma_{5})T^{a}_{\alpha \beta} b_{\beta}]
\;G^{a}_{\mu \nu}.
\end{equation}
$\alpha$ and $\beta$ are color indices, the subscripts V$\mp $A
represent the chiral projections $1\mp \gamma_{5}$ and $G^{a}_{\mu \nu}$
is the gluonic field strength tensor.

In addition to the SM operators, the MSSM gluino penguin processes introduce
operators of opposite chirality to the $O_{k}$:
\begin{equation}
\label{bqcd19d}
\tilde{O}_{P}=\tilde{O}_{4}+\tilde{O}_{6}-
\frac{1}{3}(\tilde{O}_{3} +\tilde{O}_{5}),
\end{equation}
where
\begin{eqnarray}
\label{bqcd22}
\tilde{O}_{3,5} & \equiv & (\bar{q}_{\alpha}\gamma^{\mu}
b_{\alpha})_{\text{V+A}}\;
\sum_{q^{\prime}}(\bar{q}^{\prime}_{\beta}\gamma_{\mu}
q^{\prime}_{\beta})_{\text{V} \pm \text{A}}  \\
\tilde{O}_{4,6}& \equiv & (\bar{q}_{\alpha}\gamma^{\mu}
b_{\beta})_{\text{V+A}}\;
\sum_{q^{\prime}}(\bar{q}^{\prime}_{\beta}\gamma_{\mu}
q^{\prime}_{\alpha})_{\text{V} \pm \text{A}},  
\end{eqnarray}
and 
\begin{equation}
\label{bqcd23}
\tilde{O}_{8}\equiv \frac{g_{s}^{2}}{8 \pi ^{2}}m_{b}[\bar{q}_{\alpha}
\sigma^{\mu \nu}(1-\gamma_{5})T^{a}_{\alpha \beta} b_{\beta}]
\;G^{a}_{\mu \nu}.
\end{equation}

The effective Hamiltonian for $\Delta B =1$ decays at scale 
$\mu =O(m_{b})$ has the structure
\begin{eqnarray}
\label{bqcd32}
H_{\text{eff}}(\mu )& = & \frac{G_{F}}{\sqrt{2}}\{V^{*}_{uq}V_{ub}
[C_{1}(\mu )O^{u}_{1}+C_{2}(\mu )O^{u}_{2}] 
 +V^{*}_{cq}V_{cb}
[C_{1}(\mu )O^{c}_{1}+C_{2}(\mu )O^{c}_{2}]  \nonumber  \\
&& -V^{*}_{tq}V_{tb}\sum_{k=3, \ldots ,6,8}\;
[C_{k}(\mu )O_{k}+\tilde{C}_{k}(\mu )\tilde{O}_{k}]\}
\end{eqnarray}
where
\begin{eqnarray}
\label{bqcd31}
O^{q^{\prime}}_{1} & \equiv & (\bar{q}_{\alpha}\;
q^{\prime}_{\alpha})_{\text{V-A}}\;
(\bar{q}^{\prime}_{\beta}\;b_{\beta})_{\text{V-A}}, \\
O^{q^{\prime}}_{2} & \equiv & (\bar{q}_{\alpha}\;
q^{\prime}_{\beta})_{\text{V-A}}\;
(\bar{q}^{\prime}_{\beta}\;b_{\alpha})_{\text{V-A}}, 
\end{eqnarray}
($q^{\prime} \in \{u,c\}$), are the tree current-current operators.
We omit the photon magnetic dipole operator $O_{7}$ and electroweak
penguin operators.
The Wilson coefficients $C_{k}(\mu )=C^{\text{SM}}_{k}$ for $k=1,2$ and 
$C_{k}(\mu )=C^{\text{SM}}_{k}+C^{\text{MSSM}}_{k},\;
\tilde{C}_{k}(\mu )=\tilde{C}^{\text{MSSM}}_{k}$ for $k=3, \ldots ,6,8$
incorporate the physics contributions from scales higher than $\mu $ and
are determined perturbatively at $M_{W}$ by matching to the full 
SM+MSSM theory.  
Noting that the $u$ and $c$ contributions from the SM penguin to the
coefficient of $O_{P}$ in (\ref{bqcd20}) cancel in this matching 
process \cite{Buras99}, we obtain (in
the naive dimensional regularization (NDR) scheme)
\begin{eqnarray}
\label{bqcd33}
C^{\text{SM}}_{1}(M_{W})&=&1-\frac{11}{6}\; \frac{\alpha_{s}(M_{W})}{4 \pi}, \\
C^{\text{SM}}_{2}(M_{W})&=& \frac{11}{2}\; \frac{\alpha_{s}(M_{W})}{4 \pi},   \\
C^{\text{SM}}_{k}(M_{W})&=& \frac{\alpha_{s}(M_{W})}{24 \pi}\;A_{k}
\;[f_{1}(x_{t})-\frac{2}{3}], \\
C^{\text{SM}}_{8}(M_{W}) &=& \frac{1}{2}\;f_{2}(x_{t}), \\
\label{bmssm}
C^{\text{MSSM}}_{k}(M_{W})&=& \frac{\alpha_{s}(M_{W})}
{24 \pi}\;(\frac{G_{F}}{\sqrt{2}}V^{*}_{tq}V_{tb})^{-1}\;
A_{k}F^{L}_{1}(0),  \\
\label{bmssm2}
C^{\text{MSSM}}_{8}(M_{W})&=& \frac{1}{2}\;
(\frac{G_{F}}{\sqrt{2}}V^{*}_{tq}V_{tb})^{-1}\;F^{L}_{2}(0) 
\end{eqnarray}
where $k=3, \ldots ,6$ and
\begin{equation}
\label{bqcd25}
A_{k}\equiv (-1,3,-1,3).
\end{equation}
The coefficients $\{\tilde{C}^{\text{MSSM}}_{k},\tilde{C}^{\text{MSSM}}_{8}
\}$ are obtained from $\{C^{\text{MSSM}}_{k},C^{\text{MSSM}}_{8}\}$ by the
replacement $F^{L}_{1,2}(0) \rightarrow F^{R}_{1,2}(0)$.

The Wilson coefficients
\begin{equation}
\label{bqcd34}
\bbox{C}(M_{W}) \equiv (C_{1}(M_{W}), \ldots ,C_{6}(M_{W}))^{\text{T}}
\end{equation}
evolve under the renormalization group equations (RGE) to
\begin{equation}
\label{bqcd35}
\bbox{C}(\mu ) = \bbox{U}_{5}(\mu , M_{W})\bbox{C}(M_{W})
\end{equation}
where $\bbox{U}_{5}(\mu ,M_{W})$ is the five-flavor $6 \times 6$
evolution matrix. In next-to-leading-order (NLO), $\bbox{U}_{5}(\mu ,
M_{W})$ is given by \cite{Buras92,Buch96}
\begin{equation}
\label{bqcd36}
\bbox{U}_{5}(\mu ,M_{W})=[1+\frac{\alpha_{s}(\mu )}{4 \pi}\bbox{J}]
\bbox{U}^{0}_{5}(\mu , M_{W})[1-\frac{\alpha_{s}(M_{W})}{4 \pi}\bbox{J}]
\end{equation}
where $\bbox{U}^{0}_{5}(\mu ,M_{W})$ is the leading order (LO) evolution
matrix and $\bbox{J}$ expresses the NLO corrections to the evolution.
NLO Wilson coefficients in the NDR renormalization scheme were computed
from (\ref{bqcd35}) and (\ref{bqcd36}) using expressions given in
\cite{Buras92,Buch96} for $\bbox{J}$ and $\bbox{U}^{0}_{5}$ and
the two-loop expression for $\alpha_{s}(\mu )$ with
five flavors and $\alpha_{s}^{\overline{MS}}(M_{Z})=0.118 \;
(\Lambda_{\text{QCD}}=0.225$ GeV). 

The renormalization group induced mixing between $O_{8}$ and the set
$\{O_{1}, \ldots , O_{6}\}$ vanishes at 1-loop order and it is sufficient
to use the LO evolution for $C_{8}(\mu )$:
\begin{equation}
\label{bqcd37}
C_{8}(\mu ) = \eta ^{-14/23}\;C_{8}(M_{W})+\sum_{i=1}^{8}\bar{h}_{i}
\eta^{-a_{i}}
\end{equation}
where $\eta \equiv \alpha_{s}(\mu )/\alpha_{s}(M_{W})$ and the constants
$\bar{h}_{i}$ and $a_{i}$ are tabulated in \cite{Buch96}.

The MSSM generated QCD penguin operators $\{ \tilde{O}_{3}(M_{W}), \ldots ,
\tilde{O}_{6}(M_{W})\}$ and $\tilde{O}_{8}(M_{W})$ renormalize in the 
same manner as $\{O_{3}(M_{W}), \dots , O_{6}(M_{W})\}$ and $O_{8}(M_{W})$
\cite{Barb97}.

The Wilson coefficients $C_{k}(\mu ), k=1, \ldots ,6,8$, for the SM and
the combined SM+MSSM, together with the Wilson coefficients $\tilde{C}_{k}
(\mu ), k=3, \ldots , 6$, for the MSSM are listed in Table \ref{table1} 
for the scales $\mu =m_{b}$ and $\mu =2.5$ GeV.  Inclusion of the MSSM 
processes generates complex Wilson coefficients due to the complex mixing
coefficients  $\Lambda^{bq}_{ABj}$ in the MSSM form factors. These mixing 
coefficients arise from the mismatch between the interaction and 
$\tilde{d}$ mass eigenstates at the $d-\tilde{d}-\tilde{g}$ vertices and
contain non-removable RGE-generated phases in the off-diagonal components
of the couplings triggered, in particular, by the complexity of the 
large $t$-quark Yukawa coupling. The numerical values of
$\Lambda^{bq}_{ABj}$ used in (\ref{bmssm}) and (\ref{bmssm2})  
for $C^{\text{MSSM}}_{k}(M_{W}),k=3, \ldots ,6,8$
are generated from (\ref{bqcd14}) for a given MSSM data set by numerical
diagonalization of the $\tilde{d}_{j}$ mass$^{2}$ matrices at the scale
$M_{W}$.

The MSSM results are for the data set 
\begin{equation}
\label{bqcd37a}
A=-300\; \text{GeV},\quad  \mu > 0,\quad \tan \beta =48,\quad 
m_{0}=275\; \text{GeV}, \quad m_{1/2}=150 \;\text{GeV}
\end{equation}
which maximizes the effects of SUSY \cite{ACW98} as measured by the
ratio $\Gamma^{\text{Peng}}(\text{MSSM})/\Gamma^{\text{Peng}}(\text{SM})$ 
of integrated decay rates at scale $M_{W}$ for 
$b \rightarrow q q^{\prime} \bar{q}^{\prime}$. 
This ratio is largest for high $\tan \beta $ and low $(m_{0},m_{1/2}) $ 
and for the set (\ref{bqcd37a}) is  $\approx 0.10 $ for 
$b \rightarrow d$ and $\approx 0.085$ for $b \rightarrow s$. 
The data set (\ref{bqcd37a}) satisfies the constraints imposed \cite{cho99}
on the SUSY parameter space by recent experimental bounds on the mass
of the lightest chargino ($m_{\tilde{\chi }^{\pm }_{1}}> 91$ GeV) and 
lighter $\tilde{t}$ squark ($m_{\tilde{t}_{1}}>80$ GeV) and by precision
electroweak data constraints on SUSY corrections to electroweak parameters
but is on the verge of being excluded by the latest measurements for 
$\text{Br}(B \rightarrow X_{s}\gamma )$.

As expected from the observations above on the relative magnitudes of the
SM and MSSM form factors at scale $M_{W}$, the only significant MSSM
effects occur in the magnetic dipole coefficients. 
For the data set (\ref{bqcd37a}) we find $C^{\text{MSSM}}_{8}(M_{W})
=-0.4631+0.0224i\;(b \rightarrow d)$ and $-0.4755-0.0013i\;(b 
\rightarrow s)$ which are much larger than $C^{\text{SM}}_{8}(M_{W})
=-0.0953$ and comparable to the largest SM coefficient 
$C^{\text{SM}}_{1}(M_{W})=0.9828$. Thus $C_{8}(\mu )$ is greatly enhanced 
by the MSSM contribution at $M_{W}$. 
Although $\tilde{C}^{\text{MSSM}}_{8}(M_{W})$ 
is only of comparable magnitude to $C^{\text{SM}}_{3, \ldots, 6}$, it is
also enhanced by (\ref{bqcd37}) and becomes significant.

The Wilson coefficients in NLO precision depend upon the renormalization
scheme \cite{Buras92,Kramer94,Ali98} and this unphysical dependence
is compensated by the perturbative one-loop QCD corrections to the matrix
elements of the four-quark operators $O_{1}, \ldots , O_{6}$ (and 
$\tilde{O}_{3}, \ldots , \tilde{O}_{6}$) at the scale $\mu $. This is
equivalent to using the effective Hamiltonian (\ref{bqcd32}) with the 
Wilson coefficients $C_{k}(\mu ), k=1, \ldots , 6$, replaced by the 
effective $q^{2}$-dependent Wilson coefficients 
$C^{\text{eff}}_{k}(\mu ,q^{2})$ such that
\begin{equation}
\label{bqcd37b}
C_{k}(\mu )\langle O_{k}(\mu ) \rangle = C_{k}^{\text{eff}}(\mu ,q^{2})
\langle O_{k} \rangle ^{\text{tree}}.
\end{equation}
The two sets of Wilson coefficients are related by \cite{Ali98} (in the NDR
scheme)
\begin{eqnarray}
\label{bqcd38}
C^{\text{eff}}_{k}(\mu , q^{2}) & =& [1+\frac{\alpha_{s}(\mu )}{4 \pi}
(r^{\text{T}}_{V}+\gamma^{\text{T}}_{V}\log \frac{m_{b}}{\mu }]_{kj}
\;C_{j}(\mu ) \nonumber \\
&& +\frac{\alpha_{s}(\mu )}{24 \pi }A^{\prime}_{k}(c_{t}+c_{p}+c_{g})
\end{eqnarray}
where $A^{\prime}_{k}\equiv (0,0,-1,3,-1,3)^{\text{T}}$, the $6\times 6$
constant matrices $r_{V}$ and $\gamma_{V}$ (given in \cite{Ali98}) represent
the process-independent parts of the vertex corrections and $c_{t},\;c_{p}$
and $c_{g}$ are the contributions arising from the penguin-type corrections
to the current-current operators $O_{1,2}$, penguin-type corrections to
$O_{3}, \ldots , O_{6}$, and the tree level diagram of $O_{8}$ respectively.
The quantities $c_{t}, \;c_{p}$ and $c_{g}$ are given by \cite{Ali98}
\begin{eqnarray}
\label{bqcd39}
c_{t} & = & -\frac{C_{1}}{V^{*}_{tq}V_{tb}}[V^{*}_{uq}V_{ub}(\frac{2}{3}
+{\cal F}_{1}(m_{u},q^{2})) 
 + V^{*}_{cq}V_{cb}(\frac{2}{3}+{\cal F}_{1}(m_{c},q^{2}))],  \\
\label{bqcd40}
c_{p} & = & C_{3}[\frac{4}{3} + {\cal F}_{1}(m_{q},q^{2})+
{\cal F}_{1}(m_{b},q^{2})] 
 + (C_{4}+C_{6})\sum_{j=u,d,s,c,b} {\cal F}_{1}(m_{j},q^{2}), \\
\label{bqcd41}
c_{g} & = & -\frac{2 m_{b}}{\sqrt{q^{2}}}\;C_{8}
\end{eqnarray}
where
\begin{equation}
\label{bqcd42}
{\cal F}_{1}(m,q^{2}) \equiv \frac{2}{3} \log (\frac{m^{2}}{\mu ^{2}})
- \Delta F_{1}(\frac{q^{2}}{m^{2}})
\end{equation}
and
\begin{equation}
\label{bqcd43}
\Delta F_{1}(z) = -4 \int^{1}_{0}\;dx x(1-x)\log[1-zx(1-x)-i\epsilon].
\end{equation}
Note that ${\cal F}_{1}(m,q^{2})$ is just $-f_{1}(x,q^{2})$ of \cite{ACW98}
with $M_{W}^{2}$ replaced by $\mu ^{2}$ and is complex for $q^{2} > 4 m^{2}$.
These strong phases in ${\cal F}_{1}$, generated at the $u\bar{u}$ and 
$c\bar{c}$ thresholds, combine with the weak CKM phases to produce 
effective Wilson coefficients $C^{\text{eff}}_{k}, k=3, \ldots ,6$, which
differ in both magnitude and phase for $b$ and $\bar{b}$ decays and hence 
CP violation is manifest.

Some comments are appropriate about whether or not to include 
the term $c_{g}$ in
(\ref{bqcd38}). This term is absent in most applications of the 
effective Hamiltonian method \cite{Fleisch93,Desh98,Cheng98} and was first
included by \cite{Ali98}.
The $O_{8}$ contribution to $\langle q q^{\prime} 
\bar{q}^{\prime}|H_{\text{eff}}|b\rangle$ can be written
\begin{equation}
\label{bqcd44}
M_{8} = i \frac{G_{F}}{\sqrt{2}}\frac{\alpha_{s}}{\pi}F^{R}_{2}(0)
\frac{m_{b}}{q^{2}}
  [\bar{u}_{q}\gamma_{\mu}\gamma_{\nu}q^{\nu}(1+\gamma_{5})
T^{a}u_{b}][\bar{u}_{q^{\prime}}\gamma^{\mu }T^{a}v_{\bar{q}^{\prime}}].
\end{equation}
In the factorization model used by  \cite{Ali98} color considerations
preclude $q^{\prime}$ and $\bar{q}^{\prime}$ combining into the same meson 
and it is reasonable to assume that 
$\bbox{p}_{q^{\prime}}=-\bbox{p}_{\bar{q}^{\prime}}$ 
in the $b$ rest frame. The gluon momentum
is  then
\begin{equation}
\label{bqcd44a}
q^{\mu}=\sqrt{q^{2}}\;p^{\mu }_{b}/m_{b}. 
\end{equation}
This allows $M_{8}$ to be expressed in terms of the penguin operator 
combination $O_{P}$ given in (\ref{bqcd19c}), and yields the contribution
$c_{g}$ to $C^{\text{eff}}_{k}, k=3, \ldots , 6$. 
$q^{2}$ is conventionally
replaced by an averaged value $\langle q^{2}\rangle $ taken to lie in 
the range 
$m_{b}^{2}/4 \leq \langle q^{2}\rangle \leq m_{b}^{2}/2 $.
If the assumption (\ref{bqcd44a}) is not valid then $c_{g}$ should be
omitted and a proper treatment of 
the magnetic dipole contribution to decay rates must be based upon 
(\ref{bqcd44}). This then involves an integration over a range of $q^{2}$ 
determined by kinematical considerations (see later and Refs. 
\cite{Desh96,ACW98,Hou99}).

We have computed effective Wilson coefficients with and without the
term $c_{g}$. Although the assumption
$\bbox{p}_{q^{\prime}}=-\bbox{p}_{\bar{q}^{\prime}}$ 
is valid for many of the processes (so-called {\it A processes}) considered
in our spectator model (see later), we choose not to make this assumption
in our effective Wilson coefficients and hence do not include the $c_{g}$
term in our calculations.

We list in Table \ref{table3} the effective Wilson coefficients $C^{\text{eff}}_{k}(
\mu , q^{2}), k=1, \ldots , 6$ and $\tilde{C}^{\text{eff}}_{k}(\mu ,q^{2}),
k=3, \ldots ,6$ evaluated from (\ref{bqcd38}) 
at $q^{2}=m_{b}^{2}/2$ and $\mu = $ 2.50 GeV for
the processes $b \rightarrow d q \bar{q}^{\prime}$ and $\bar{b} 
\rightarrow \bar{d} q \bar{q}^{\prime}$. Table \ref{table4} 
lists the same quantities 
for the processes $b \rightarrow s q \bar{q}^{\prime}$ and $\bar{b} 
\rightarrow \bar{s} q \bar{q}^{\prime}$.  
The values of the $\tilde{C}^{\text{eff}}_{k}$ coefficients are due nearly
entirely to the $A_{k}^{\prime}$ term in (\ref{bqcd38}). 
If the input CKM matrix is changed from the 
Particle Data Group parameterization to the Wolfenstein form with the
parameter values $A=0.81,\;\rho = 0.05$ and $\eta =0.36$ used by 
\cite{Ali98}, and we include $c_{g}$, 
we obtain values for the SM $C^{\text{eff}}_{k}$ which
agree very closely with those given in \cite{Ali98}.

\section{Decay rates for \lowercase{
$b \rightarrow q q^{\prime} \bar{q}^{\prime\prime}$}}

In calculating the $b$ quark decay rates we allow for the more general case
where the antiquark produced is not necessarily $\bar{q}^{\prime}$. 
We firstly consider the SM only and, in accordance with most 
studies of the SM, 
omit the small magnetic dipole term $C^{\text{SM}}_{8}$ 
from the effective Hamiltonian (\ref{bqcd32}) in the calculation of
decay rates.

The partial decay rate, $b$ spin averaged and summed over final spin
states, has overall spherical symmetry. Apart from its overall orientation,
a final state is specified by only two parameters, say 
$p_{q}=|\bbox{p}_{q}|$ and $p_{q^{\prime}}=|\bbox{p}_{q^{\prime}}|$.
The partial decay rate in the $b$ rest frame is
\begin{equation}
\label{bqcd48}
\frac{d^{2}\Gamma}{dp_{q}dp_{q^{\prime}}}=\frac{G_{F}^{2}}{\pi^{3}}p_{q}
p_{q^{\prime}}E_{\bar{q}^{\prime\prime}}[\alpha_{1}\; \frac{p_{q}
\cdot p_{q^{\prime}}}{E_{q}E_{q^{\prime}}}+\alpha_{2}\;
\frac{p_{q}\cdot p_{\bar{q}^{\prime\prime}}}
{E_{q}E_{\bar{q}^{\prime\prime}}} 
 +\alpha_{3}\;\frac{m_{q^{\prime}}m_{\bar{q}^{\prime\prime}}}
{E_{q^{\prime}}E_{\bar{q}^{\prime\prime}}}]
\end{equation}
where
\begin{eqnarray}
\label{bqcd46}
\alpha_{1} & = & |d_{1}+d_{2}+d_{3}+d_{4}|^{2}+2|d_{1}+d_{4}|^{2}
+2|d_{2}+d_{3}|^{2}, \nonumber  \\
\alpha_{2} & = & |d_{5}+d_{6}|^{2}+2|d_{5}|^{2}+2|d_{6}|^{2},\nonumber \\
\alpha_{3} & = & \Re[(3d_{1}+d_{2}+d_{3}+3d_{4})d_{6}^{*} 
 +(d_{1}+3d_{2}+3d_{3}+d_{4})d_{5}^{*}]
\end{eqnarray}
and
\begin{equation}
\label{bqcd47}
d_{1,2}(\mu ,q^{2})  \equiv  V_{q^{\prime \prime}q}^{*}V_{q^{\prime}b}\;
C^{\text{eff}}_{1,2}(\mu ,q^{2}), \quad
d_{3, \ldots ,6}(\mu ,q^{2})  \equiv  -V_{tq}^{*}V_{tb}\;
C^{\text{eff}}_{3, \ldots ,6}(\mu ,q^{2}).
\end{equation}
Here $p_{q}\cdot p_{q^{\prime}}=[m_{b}^{2}+m_{\bar{q}^{\prime\prime}}^{2}-
m_{q}^{2}-m_{q^{\prime}}^{2}-2m_{b}E_{\bar{q}^{\prime\prime}}]/2 $,
$p_{q}\cdot p_{\bar{q}^{\prime\prime}}=[m_{b}^{2}+m_{q^{\prime}}^{2}
-m_{q}^{2}-m_{\bar{q}^{\prime\prime}}^{2}-2m_{b}E_{q^{\prime}}]/2$ 
and the
angles between the particle velocities must be physical, for example
$|\cos (\theta_{qq^{\prime}})| \leq 1$.
Note that, for $q\in\{d,s\}$, only the tree current-current terms $d_{1,2}$
contribute when $q^{\prime} \neq q^{\prime\prime} \in \{u,c\}$ whereas, 
for $\Delta C=\Delta U=0$ transitions, only
the penguin terms $d_{3, \ldots , 6}$ contribute for 
$q^{\prime} = q^{\prime\prime} \in \{d,s\}$ and both tree and penguin terms
contribute for $q^{\prime} = q^{\prime\prime} \in \{u,c\}$. 

For massless final state quarks, the decay rate is \cite{Fleisch93}
\begin{equation}
\label{bqcd49}
\Gamma = 12 \Gamma_{0}\int^{1}_{0} d\xi [\alpha_{1}(\xi)+\alpha_{2}(\xi)]
[\frac{1}{6}-\frac{\xi^{2}}{2}+\frac{\xi^{3}}{3}]
\end{equation}
where $\Gamma_{0}\equiv G_{F}^{2}m_{b}^{5}/192 \pi^{3}$ and $\xi \equiv 
q^{2}/m_{b}^{2}$. For SM pure penguin decays, the QCD corrected decay 
rates at scale $\mu $ calculated from (\ref{bqcd49}) generally agree to 
better than 30\% (see Table \ref{table5}) with 
$\eta ^{2}\Gamma^{(0)}$ where $\Gamma^{(0)}$ are the weak scale 
decay rates calculated using $q^{2}$-dependent form factors \cite{ACW98} 
and the factor $\eta ^{2}$ 
accounts for the running of $\alpha_{s}$ 
from $M_{W}$ to $\mu $. To this extent the results support the 
conjecture of G\'{e}rard and Hou \cite{Gerard91} mentioned in the 
Introduction.

To study the effects of the MSSM on the quark decay rates, we note 
that the only significant effects of the MSSM are confined to 
the magnetic dipole coefficients $C_{8}(\mu )$ and $\tilde{C}_{8}(\mu )$ 
(see Table \ref{table1}) and the terms 
$\tilde{C}_{k}, k=3, \ldots ,6$ of the effective
Hamiltonian (\ref{bqcd32}) can be neglected in any calculation of decay 
rates.  
Since $|\tilde{C}_{8}(\mu )|/|C_{8}(\mu )|\lesssim 0.2$ for both
$b \rightarrow d$ and $b \rightarrow s$ transitions, it would be reasonable
to also neglect the $\tilde{C}_{8}$ term so that, if the assumption 
(\ref{bqcd44a}) about the momenta of $q^{\prime}$ and
$\bar{q}^{\prime\prime}$ is applicable, the major effects of the MSSM
could be incorporated in a modified value of $c_{g}$. However, for the
general case, the $c_{g}$ term should be omitted from $C^{\text{eff}}_{k},
\;k=3, \ldots ,6 $ and the magnetic dipole terms directly included.
For massless final state quarks the dipole contributions add the terms 
\cite{ACW98}
\begin{equation}
\label{bqcd49a}
\frac{1}{3}(|d_{8}|^{2}+|\tilde{d}_{8}|^{2})(\frac{1}{3\xi}-\frac{1}{2}
+\frac{\xi^{2}}{6})  
-\frac{2}{3}\;\Re[(d_{1}+d_{4}+d_{6})d_{8}^{*}](1-\xi)^{2}
\end{equation}
to the integrand of (\ref{bqcd49}), where
\begin{equation}
\label{bqcd49b}
d_{8}(\mu )\equiv -V^{*}_{tq}V_{tb}\; C_{8}(\mu ), \quad 
\tilde{d}_{8}(\mu )\equiv -V^{*}_{tq}V_{tb} \;\tilde{C}_{8}(\mu ), 
\end{equation}
thus requiring the imposition of a lower cutoff on $q^{2} $ 
(taken to be 1.0 GeV$^{2}$).  The MSSM gluino penguin processes so included 
are greatly
enhanced relative to the SM penguin amplitudes by the 
renormalization-induced QCD corrections (see Table \ref{table5}) and the
resulting decay rates $\Gamma $ for the SM+MSSM far exceed the 
$\eta ^{2}$-scaled weak scale decay rates $\Gamma^{0}$.
We also note from Table \ref{table5} that, whereas the QCD corrections
only mildly affects the magnitudes of the $CP$ asymmetries $a_{CP} 
\equiv (\Gamma -\bar{\Gamma })/(\Gamma + \bar{\Gamma})$ for the SM, they can
decrease the magnitude and change the sign of $a_{CP}$ for the SM+MSSM.

The impact on $b \rightarrow sg$ of new physics exhibited by an enhanced
chromomagnetic coupling $C_{8}$ involving an unconstrained new CP phase
has recently been studied by Hou \cite{Hou99} within an effective
Hamiltonian framework. $q^{2}$-dependent decay rates and CP asymmetries
are given for phase differences of $\pi /4,\;\pi /2$ and $3\pi /4$ 
between the $C_{8}$ amplitude and the complex SM $C_{3, \ldots ,6}$
penguin amplitudes. In all cases Hou finds large rate asymmetries above
the $c\bar{c}$ threshold, a feature not present in our present 
calculations, or in our earlier weak scale calculations incorporating
a $q^{2}$-dependent SM $F^{L}_{1}(q^{2})$ form factor \cite{ACW98}, because
of the much smaller MSSM phases (for $b \rightarrow s$ the phases of the
dominant MSSM form factors $F^{R}_{2}$ and $F^{L}_{1}$ are  
nearly constant \cite{ACW98} at 
$\approx -0.016$ over the allowed MSSM parameter space).

Although the QCD corrected SM+MSSM decay rates are quite sensitive 
to the choice of lower cutoff for $q^{2}$, it is clear that the values
obtained for the MSSM data set (\ref{bqcd37a}) are unacceptably large.
For example, assuming a maximum branching ratio of 1\% for $b \rightarrow
sd\bar{d}$ gives an experimental upper bound on the $b \rightarrow s$
decay rate of $\lesssim 3 \times 10^{-15}$ GeV, a factor of more than 
10 smaller
than our calculated rate. Even though the MSSM data set used in the 
present calculations
is not typical of the MSSM parameter space, having been chosen to maximize
the hadronic $b$ decay rate at $M_{W}$, our findings here suggest that
experimental data on hadronic $b$ decays will exclude a similar low 
$(m_{0},m_{1/2})$ region of the MSSM parameter space to that excluded 
by the $b \rightarrow s\gamma $ constraints \cite{cho99}.

From here on we focus on decay rates in the SM for $b$ decays which  
are not pure penguin processes. We therefore neglect the 
magnetic dipole  
interaction and retain final state quark masses. If
we make the common assumption that the coefficients $\alpha_{i}, i=1,2,3$ 
are constant, 
evaluated at $q^{2}=m_{b}^{2}/2$ for example, then the decay rate is
\begin{equation}
\label{bqcd50}
\Gamma = \Gamma_{0}[\alpha_{1}(\xi =\case{1}{2})I_{1}
+\alpha_{2}(\xi = \case{1}{2})I_{2}+\alpha_{3}(\xi =\case{1}{2})I_{3}]
\end{equation}
where the phase space integrals $I_{i},i=1,2,3$ have to be 
computed numerically. Decay rates computed from (\ref{bqcd50}) are shown
in Table \ref{table6}, together with branching 
ratios for various $b$ decays. Branching ratios  
are less sensitive to the value chosen
for $m_{b}$ and have been
obtained by dividing the decay rates by the total decay rate
$\Gamma^{\text{Total}}$ computed using the $O(\alpha_{s})$ QCD-corrected 
semileptonic decay rate
\begin{equation}
\label{bqcd50a}
\Gamma (b \rightarrow cl^{-}\bar{\nu}_{l}) = \Gamma_{0}|V_{cb}|^{2}I_{1}\;
[1-\frac{2 \alpha_{s}}{3 \pi} f(\rho_{c}) ],
\end{equation}
where $\rho_{c} \equiv (m_{c}/m_{b})^{2}$, with $f(\rho_{c})=2.51 $
\cite{Nir89}.

\section{A spectator quark model for $B \rightarrow \lowercase{h_{1}h_{2}}$}

We now wish to make the transition from decay rates at the quark level to
decay rates for two body hadronic decays $B \rightarrow h_{1}h_{2}$. 
Calculation of the required hadronic matrix elements 
$\langle h_{1}h_{2}|O_{k}|B\rangle $ from first principles is currently
not possible and several approximate schemes are available in the literature.
Much work has been done on the factorization model \cite{fact} 
and the more recent generalised factorization model 
\cite{Ali98,Cheng98,Ali98a} of this process in which final state 
interactions are
neglected and the hadronic matrix elements are factorized into a product of
two hadronic currents:$\langle h_{1}|J_{1\;\mu}|B \rangle \langle h_{2}| 
J^{\mu }_{2}|0\rangle $.
The operators $O^{q}_{2}$ and $O_{4,6}$ are Fierz transformed into a 
combination of color singlet-singlet
and octet-octet terms and the octet-octet terms then discarded. 
The singlet-singlet current matrix elements are then
expressed in terms of known decay rates and form factors. 
Consequently, the hadronic matrix elements are expressed in terms of 
the combinations
\begin{equation}
\label{bqcd51a}
a_{2k-1}=C^{\text{eff}}_{2k-1}+\frac{1}{N_{c}}C^{\text{eff}}_{2k}, \quad
a_{2k}=C^{\text{eff}}_{2k}+\frac{1}{N_{c}}C^{\text{eff}}_{2k-1}
\end{equation}
where $k=1,2,3$ and the number of colors $N_{c}$ is usually treated as a free 
phenomenological parameter in order to compensate for the 
discarded octet-octet terms.
Factorization works reasonably well for $B$ decays to heavy hadrons 
\cite{NeubStech98}, in that $a_{1}$ and $a_{2}$ seem to assume universal
values, and it is argued \cite{Ali98} to also account for
$B$ decays to light hadrons such as $B \rightarrow K\pi$ and
$B \rightarrow \pi \pi$. 
The generalised factorization approach has been criticised 
\cite{Buras98a,Buras99} because the effective Wilson coefficients
(\ref{bqcd38}) are gauge dependent, implying that the value of $N_{c}$  
extracted from comparisons of factorization model predictions with data
cannot have any physical meaning\footnote{Gauge invariant and infrared 
finite effective Wilson coefficients have recently become available 
\cite{infrawilson}}. These authors \cite{Buras98a} present
an alternative model based upon Wick contractions in the matrix
elements of the NLO effective Hamiltonian. Furthermore, the 
applicability of factorization to 
the hadronization of outgoing light quark pairs moving in opposite
directions has been questioned 
\cite{Bjorken97} and it has been emphasised there that hadronic $B$ decays,
especially as they are originated by three partons, are phase space driven.

In the spectator model \cite{spect} alternative to factorization, 
the $b$ quark and
spectator quark  are treated as quasifree particles with a 
distribution of momenta due to their Fermi motion relative to the $B$ meson.
In these models the $b$ quark is treated as a virtual particle with an
invariant mass $m_{b}(p_{s})$ satisfying $m^{2}_{b}(p_{s})=m_{B}^{2}
+m_{s}^{2}-2m_{B}\sqrt{p_{s}^{2}+m_{s}^{2}}$.
We, however, consider here a tentative spectator model based upon the idea of 
duality between quark and
hadron physics at the high energies of $b$ quark and $B$ meson decays. The
decays at the quark level, even including the penguins, are 
basically short distance processes. In our proposed spectator quark model
the long distance hadronization is largely a matter of incoherently 
assigning regions of the final quark phase space to the different
mesonic systems.

For example, we consider a $B$ meson $b\bar{u}$ or $b\bar{d}$ 
to be a heavy stationary
$b$ quark accompanied by a light spectator constituent antiquark 
which has a spherically symmetric normalized momentum distribution 
$ P(|\bbox{p}_{s}|^{2}) d^{3}p_{s}$. The total meson decay rate 
through a particular mode is then assumed
to be
\begin{equation}
\label{bqcd51}
\Gamma = \int\;\frac{d^{2}\Gamma}{dp_{q}dp_{q^{\prime}}}\;
P(|\bbox{p}_{s}|^{2})d^{3}p_{s}\;dp_{q}\;dp_{q^{\prime}},
\end{equation}
equal to the initiating decay rate.
Neglecting, for the moment, any constraints due to quark color, 
we suppose the spectator antiquark to combine with the quark 
$i (i=q^{\prime} $ or  $q)$ to form the meson system. 
For example, if $i=q^{\prime}$ we assign  
a mass $M_{q^{\prime}s}$ to the system such that
\begin{equation}
\label{bqcd52}
M_{q^{\prime}s}^{2} = (p_{q^{\prime}}+p_{s})\cdot (p_{q^{\prime}}+p_{s})
 =m_{q^{\prime}}^{2}+m_{s}^{2}
+2(E_{q^{\prime}}E_{s}-p_{q^{\prime}}p_{s}\cos \theta_{q^{\prime}s}).
\end{equation}
Constraining $p_{q^{\prime}}$ and $p_{s}$ to have mass $M_{q^{\prime}s}$, 
we can infer from
(\ref{bqcd51}) that
\begin{equation}
\label{bqcd53}
\frac{d\Gamma}{dM_{q^{\prime}s}} = 2M_{q^{\prime}s} \int \;
\frac{d^{2}\Gamma}{dp_{q}dp_{q^{\prime}}}\;P(|\bbox{p}_{s}|^{2})
\delta (M_{q^{\prime}s}^{2}-m_{q^{\prime}}^{2}-m_{s}^{2}
 -2(E_{q^{\prime}}E_{s}-p_{q^{\prime}}p_{s}\cos \theta_{q^{\prime}s} ))
d^{3}p_{s}\;dp_{q}\;dp_{q^{\prime}}.
\end{equation}
Hence
\begin{equation}
\label{bqcd54}
\frac{d\Gamma }{dM_{q^{\prime}s}}=2\pi M_{q^{\prime}s}\int \;
\frac{p_{s}dp_{s}}{p_{q^{\prime}}}\;
\frac{d^{2}\Gamma}{dp_{q}dp_{q^{\prime}}} P(|\bbox{p}_{s}|^{2})
dp_{q}\;dp_{q^{\prime}}
\end{equation}
where the integration region is restricted by the condition 
$|\cos \theta_{q^{\prime}s}|\leq 1$. 

We also assign a mass 
$M_{q\bar{q}^{\prime\prime}}$ to the second quark antiquark system 
such that
\begin{equation}
\label{bqcd55}
M_{q\bar{q}^{\prime\prime}}^{2} = (p_{q}+p_{\bar{q}^{\prime\prime}})
\cdot (p_{q}+p_{\bar{q}^{\prime\prime}}) 
 =(p_{b}-p_{q^{\prime}})\cdot (p_{b}-p_{q^{\prime}}) 
 =m_{b}^{2}+m_{q^{\prime}}^{2}-2m_{b}E_{q^{\prime}}.
\end{equation}
The variable $E_{q^{\prime}}$, and hence $p_{q^{\prime}}$, 
determines the mass $M_{q\bar{q}^{\prime\prime}}$. 
Taking this mass to be the 
independent variable, we have
\begin{equation}
\label{bqcd56}
\frac{d^{2}\Gamma }{dM_{q^{\prime}s}dM_{q\bar{q}^{\prime\prime}}}
 = \frac{2\pi M_{q^{\prime}s}M_{q\bar{q}^{\prime\prime}}}{m_{b}}
 \int \frac{E_{q^{\prime}}p_{s}}{p_{q^{\prime}}^{2}}\,
\frac{d^{2}\Gamma}{dp_{q}dp_{q^{\prime}}} P(|\bbox{p}_{s}|^{2})
dp_{s}\,dp_{q}.
\end{equation}
We call this mode of quark and antiquark combination {\em process A}.
Finally, by integration, we compute the partial decay rates 
$\Gamma(M_{q^{\prime}s},M_{q\bar{q}^{\prime\prime}})$ into quark systems
with masses less than $M_{q^{\prime}s}$ and $M_{q\bar{q}^{\prime\prime}}$.
With suitable binning we equate these partial decay rates with corresponding
rates into mesons. 
To calculate an exclusive decay into a particular two meson system would
require one to follow the flow of quark spins. Although the model could
be generalized to do so, we attempt to take large enough mass bins to 
enclose all the spin combinations. When searching for CP asymmetries we
would advocate looking at more inclusive decay rates which will be 
larger and hence more readily measurable.
Of particular interest are the quark antiquark
systems forming the lowest mass 0$^{-}$ and 1$^{-}$ meson states as data 
exists for charmed quark systems  which can be used to test the spectator
model. Also, data which should exhibit matter-antimatter asymmetry is eagerly
awaited for the rare decays into light quark systems.

It is also possible that the spectator antiquark combines with the 
quark $q$, for which we get
\begin{equation}
\label{bqcd57}
\frac{d^{2}\Gamma }{dM_{qs}dM_{q^{\prime}\bar{q}^{\prime\prime}}}
 = \frac{2\pi M_{qs}M_{q^{\prime}\bar{q}^{\prime\prime}}}{m_{b}}
 \int \frac{E_{q}p_{s}}{p_{q}^{2}}\,
\frac{d^{2}\Gamma}{dp_{q}dp_{q^{\prime}}} P(|\bbox{p}_{s}|^{2})
dp_{s}\,dp_{q^{\prime}}.
\end{equation}
We call this {\em process B}. 

In some meson decays, for example $B^{-} \rightarrow 
(\bar{u}d)+(\bar{u}u)$ which is initiated by the quark decay $b \rightarrow
ud\bar{u}$, the spectator $\bar{u}$ could have combined with the $u$ or 
the $d$. However, for light meson systems such as $\pi^{-}\rho^{0}$, we
find that the different combinations come from very different regions 
of phase space of the initiating $b$ decay and we conclude that 
we are not double counting. 

Turning now to the flow of color, we examine what may be regarded as 
two extreme possibilities. In the first, called here {\em model I}, we 
take the formula (\ref{bqcd48}), (\ref{bqcd56}) and (\ref{bqcd57}) at 
face value, that is we make no attempt to follow the flow of color and
assume that all color flow is taken care of
by the gluon fields in the meson system.
In the second, called here {\em model II}, we consider the possibility
that the lowest mass meson states are only formed if the quark-antiquark 
pair are in a color singlet state. That is, we assume that the color 
disruption caused by a quark-antiquark pair in a color octet state will
result in  more complex meson systems than the lowest mass 0$^{-}$ or
1$^{-}$ states.

Projecting out the color singlet states results only in a modification 
of the coefficients $\alpha_{1,2}$ of (\ref{bqcd48}). For {\em process A}
represented by (\ref{bqcd56}) they become
\begin{eqnarray}
\label{bqcd60}
\alpha_{1} &=& 3|d_{1}+\frac{1}{3}d_{2}+\frac{1}{3}d_{3}+d_{4}|^{2} \nonumber \\
\alpha_{2} &=& 3|\frac{1}{3}d_{5}+d_{6}|^{2},
\end{eqnarray}
whereas for {\em process B} represented by (\ref{bqcd57}) they are
\begin{eqnarray}
\label{bqcd61}
\alpha_{1} &=& 3|\frac{1}{3}d_{1}+d_{2}+d_{3}+\frac{1}{3}d_{4}|^{2} \nonumber \\
\alpha_{2} &=& 3|d_{5}+\frac{1}{3}d_{6}|^{2}.
\end{eqnarray}

\section{Spectator Model Results}

Calculations have been performed for two spectator quark momentum
distributions. One is a fixed spherical bag model \cite{bag} 
in which the spectator is confined
in a spherical cavity of radius $R$ and has a wave function of the 
form
\begin{equation}
\label{bqcd62}
\phi (\bbox{r})=\frac{1}{\pi }\sqrt{\frac{\Lambda }{2}} \frac{\sin 
( \Lambda r)}{r}.
\end{equation}
This implies a momentum distribution
\begin{equation}
\label{bqcd63}
P(|\bbox{p}_{s}|^{2})=(\frac{\Lambda }{\pi })^{3}\frac
{\sin ^{2}(\pi p_{s}/\Lambda )}{|\bbox{p}_{s}|^{2}(\Lambda ^{2} -
|\bbox{p}_{s}|^{2})^{2}}
\end{equation}
and a mean square momentum 
$\overline{|\bbox{p}_{s}|^{2}}=\Lambda ^{2}$.
The parameter $\Lambda $ is determined from the mean square $B$ meson 
radius 
\begin{equation}
\label{bqcd64}
\overline{r^{2}}=\frac{1}{\Lambda ^{2}}[\frac{\pi ^{2}}{3} - \frac{1}{2}].
\end{equation}
Taking $\sqrt{\overline{r^{2}}}=0.55 $ fm yields the value $\Lambda = 
0.6 $ GeV which we have used in our calculations. As a check on the 
sensitivity of our results to the model, we have also used a Gaussian 
spectator quark distribution with
$\overline{|\bbox{p}_{s}|^{2}}=\Lambda ^{2}$:
\begin{equation}
\label{bqcd65}
P(|\bbox{p}_{s}|^{2})=(\frac{3}{2 \pi \Lambda })^{3/2}\;
e^{-3 |\bbox{p}_{s}|^{2}/2 \Lambda ^{2}}.
\end{equation}

For the maximum mass of the quark-antiquark systems we take a value midway
between the lowest mass 1$^{-}$ state and the next most massive meson.
Thus we take, for $(c\bar{u})$ or $(c\bar{d})$, $M_{c\bar{u}}=2.214$ GeV
between the $D^{*}(2.007)$ and the $D^{*}(2.420)$; for $(c\bar{s})$,
$M_{c\bar{s}}=2.323$ GeV between $D^{*}_{s}(2.112)$ and $D^{*}_{s}(2.533)$;
for $(s\bar{u})$ or $(s\bar{d})$, $M_{s\bar{u}}=1.081$ GeV between 
$K^{*}(0.892)$ and $K^{*}(1.270)$; for $(u\bar{d})$, $M_{u\bar{d}}=0.877$
GeV between $\rho(0.770)$ and $a_{0}(0.984)$; and for $(u\bar{u})$ and 
$(d\bar{d})$, $M_{u\bar{u}}=M_{d\bar{d}}=0.870$ GeV between $\omega (0.782)$
and $\eta^{\prime}(0.958)$. For the $(c\bar{c})$ system we make an 
exception and take the maximum $M_{c\bar{c}}=3.552$ GeV between 
$\chi (3.417)$ and $\psi (2s)(3.686)$.

The results are presented in Table \ref{table7} and compared, where data is available,
with the sum of the branching ratios into mesons with masses less than the 
above cutoff masses. All calculations are based upon partial decay rates
(\ref{bqcd48}) with the $\alpha_{i}$ coefficients evaluated at $q^{2}=
m_{b}^{2}/2$.
The semileptonic decay rates are also shown in Table \ref{table7}.
For these decays the lepton momenta cover all the allowed phase space.

We have repeated the calculations using the SM+MSSM effective Wilson 
coefficients of Tables \ref{table3} and \ref{table4} but neglecting the
very small $\tilde{C}^{\text{eff}}$ coefficients. The MSSM effects are
insignificant for the $b \rightarrow u d \bar{u}$ transitions. For
type {\em A } processes all branching ratios are slightly increased 
$(\sim 1\%)$. For $b \rightarrow u s \bar{u}$ transitions the MSSM effects 
are more significant and result in branching ratio decreases for type
{\em A} decays of 23\% for $b \rightarrow u s \bar{u}$ and 28\% for
$\bar{b} \rightarrow \bar{u} \bar{s} u$. The $b \rightarrow d s \bar{d}$
and $\bar{b} \rightarrow \bar{d} \bar{s} d$ branching ratios are also
decreased by about 30\% Inclusion into our spectator model of the 
MSSM-enhanced magnetic dipole term is a matter for future investigation.

\section{Discussion and conclusions}

We have investigated two aspects of hadronic $B$ decays. The first is  
the effects of renormalization-induced QCD corrections 
on the inclusion of SUSY gluino penguin contributions to the hadronic
$b$ decays $b \rightarrow qg \rightarrow q q^{\prime}\bar{q}^{\prime}$, 
$q^{\prime} \in \{d,s\}$, which have been studied  by 
calculating the Wilson coefficients of the
effective SM+MSSM Hamiltonian. The SUSY enhancement of 
these gluon mediated processes at $M_{W}$ by the MSSM magnetic dipole 
transition, which is especially significant for \cite{ACW98}
large $\tan \beta $ and low $(m_{0},m_{1/2})$, 
is further amplified by the QCD corrections, resulting in 
quark decay rates at scales $\mu \sim O(m_{b})$ for pure SM+MSSM 
penguin processes which greatly exceed those obtained by simply scaling
the rates at $M_{W}$ by the factor $\eta ^{2}$ to allow for the 
renormalization of the strong coupling constant from $M_{W}$ to $\mu $.
The rates are also significantly larger than estimated experimental
bounds, suggesting that 
experimental data on hadronic $b$ decays will exclude a similar low 
$(m_{0},m_{1/2})$ region of the MSSM parameter space to that excluded 
by the $b \rightarrow s\gamma $ constraints \cite{cho99}.
We find that, whereas the QCD corrections only mildly affect the magnitudes
of the CP asymmetries $a_{CP}$, they can change their sign.

The second aspect of hadronic $B$ decays studied is the hadronization of 
the final state quarks. We have adopted a spectator model as an alternative
to the widely used factorization models.
The semileptonic decay rates given in Table \ref{table7} 
indicate semi-quantitative
agreement between our model and data. Where there is data, the same is 
true for the hadronic decays that proceed by {\em process A}, suggesting
that color suppression does not have a large influence on these processes.
This is not so for {\em process B} hadronic decays which we find to be 
very sensitive to color suppression and there is clear evidence for color
suppression in the data. In fact, for decay modes which can proceed only
through {\em process B}, the data are for the most part upper bounds. The 
decays $B^{0}\rightarrow \psi(1s)+(K^{0},K^{*\;0})$ and $B^{+}\rightarrow
\psi (1s)+(K^{+},K^{*\;+})$, for which the experimental branching ratios are 
$(2.24 \pm 0.3)\times 10^{-3}$ and $(2.46 \pm 0.37 )\times 10^{-3}$
respectively, are {\em process B} decays in the context of our model. Taking
a cutoff mass $M_{c\bar{c}}=3.257$ GeV to include $\psi (1s)$ but exclude
the $\chi (1P)$ we find, with no color suppression, a branching ratio of
$9.22 \times 10^{-3}$. This suggests a color suppression factor of about 
four, not the orders of magnitude obtained from insisting that the $c\bar{c}$
pair are produced as a color singlet ({\em model II}).

The decays into light quark systems exhibit matter-antimatter asymmetry. 
The present calculations indicate that, averaged over a few low mass 
$\pi $ mesonic states, the $B^{+}$ and $B^{0}$ branching ratios are larger
than those for $B^{-}$ and $\bar{B}^{0}$; the opposite is true if a $K$
meson is involved. The calculations suggest the asymmetry is not large.
In the effective Hamiltonian used here, this asymmetry comes only from
the "strong" phases of the penguin diagrams. The renormalization group
construction of the effective Hamiltonian, although incorporating 
important QCD improvements, does not include the "strong" phases which 
must be present even at the first order of $\alpha_{s}$ corrections. The 
penguin includes just one $O(\alpha_{s})$ correction to the "strong" 
phase; it would be of interest to include all $O(\alpha_{s})$ corrections
to the "strong" phase.

We have compared our results for the light quark systems with the 
factorization model calculations of \cite{Kramer94} which presents 
results for pseudoscalar and 
vector meson final states. Because of the constraints of spin and isospin,
which are averaged over in our model, we are reluctant to use cuts which
bracket individual particles. However, the results of \cite{Kramer94} 
when summed over the pseudoscalar and vector final particles are generally
smaller than our calculations, but by less than factors of two. Such
factors can easily be accomodated within the uncertainties of just the 
CKM parameters for the light quark couplings.

\acknowledgments

IBW would like to thank Xin-Heng Guo for helpful discussions and to 
acknowledge support from the Center for Subatomic Structure of Matter, 
University of Adelaide, where part of this research was undertaken.

\appendix

\section*{Input parameters for numerical calculations}

We use the standard Particle Data Group \cite{pdg} parameterisation of
the CKM matrix with the central values
\begin{equation}
\label{a1}
\theta_{12}=0.221, \quad \theta_{13}=0.0035, \quad \theta_{23}=0.041
\end{equation}
and choose the CKM phase $\delta_{13}$ to be $\pi /2$.

Following Ali and Greub \cite{Ali98}, we treat internal quark masses 
in penguin loops as constituent masses with the values
\begin{equation}
\label{a2}
m_{d} = m_{u}=0.2 \;\text{GeV},\quad m_{s}=0.5 \;\text{GeV}, \quad 
m_{c} = 1.5 \;\text{GeV},\quad m_{b}=4.88 \;\text{GeV}.
\end{equation}

For the light quarks in the spectator model we take
\begin{equation}
\label{a3}
m_{u}=0.005 \;\text{GeV},\quad m_{d}=0.01\;\text{GeV}, \quad
m_{s}=0.2\;\text{GeV}
\end{equation}
and a spectator quark mass of 0.01 GeV.

\mediumtext
\begin{table}
\caption{Wilson coefficients $C_{k}(\mu )$ for the SM and the combined 
SM+MSSM and Wilson coefficients $\tilde{C}_{k}$ for the MSSM at 
the renormalization scales $\mu =m_{b}$ and $\mu =2.50 $ GeV. 
The MSSM results are for
the MSSM data set (\ref{bqcd37a}).
\label{table1}}
\begin{tabular}{ldddd}
&\multicolumn{2}{c}{SM}&\multicolumn{2}{c}{SM+MSSM} \\
$\mu $&\multicolumn{1}{c}{\phantom{xxxxx}$m_{b}$}
&\multicolumn{1}{c}{\phantom{xxxx}2.50 GeV}
&\multicolumn{1}{c}{$m_{b}$}
&\multicolumn{1}{c}{2.50 GeV}\\
\tableline
$C_{1}$&1.0767&1.1266&1.0767+0.0000i&1.1266+0.0000i \\
$C_{2}$&$-$0.1811 &$-$0.2751&$-$0.1811+0.0000i&$-$0.2751+0.0000i \\
$C_{3}$&0.0119&0.0181&0.0119+0.0000i&0.0182+0.0000i\\
$C_{4}$&$-$0.0331&$-$0.0461&$-$0.0332+0.0000i&$-$0.0462+0.0000i \\
$C_{5}$&0.0094&0.0113&0.0094+0.0000i&0.0113+0.0000i \\
$C_{6}$&$-$0.0398&$-$0.0598&$-$0.0399+0.0000i&$-$0.0599+0.0000i \\
$C_{8}$&$-$0.1449&$-$0.1585&$-$0.4768+0.0160i&$-$0.4556+0.0144i \\
$\tilde{C}_{3}$\tablenotemark[1]&&&$-$0.39($-$8)+0.19($-$9)i
&$-$0.43($-$8)+0.21($-$9)i\\
$\tilde{C}_{4}$\tablenotemark[1]&&&0.71($-$8)$-$0.34($-$9)i
&0.68($-$8)$-$0.33($-$9)i\\
$\tilde{C}_{5}$\tablenotemark[1]&&&$-$0.14($-$8)+0.65($-$10)i
&$-$0.79($-$9)+0.39($-$10)i\\
$\tilde{C}_{6}$\tablenotemark[1]&&&0.11($-$7)$-$0.55($-$9)i&
0.13($-$7)$-$0.65($-$9)i\\
$\tilde{C}_{8}$&&&$-$0.0768+0.0000i&$-$0.0976+0.0000i \\
\end{tabular}
\tablenotemark[1]{Numbers in parentheses denote powers of 10}
\end{table}

\mediumtext
\begin{table}
\caption{Effective Wilson coefficients $C_{k}^{\text{eff}}(\mu ,q^{2})$ 
for the SM and the combined 
SM+MSSM and effective Wilson coefficients $\tilde{C}_{k}^{\text{eff}}$ 
for the MSSM at 
the renormalization scale $\mu =2.50$ GeV and momentum $q^{2}=m_{b}^{2}/2$
for $b \rightarrow d$ and $\bar{b}
\rightarrow \bar{d}$ transitions. 
Results are given for the MSSM data set (\ref{bqcd37a}).
\label{table3}}
\begin{tabular}{ldddd}
&\multicolumn{2}{c}{SM}&\multicolumn{2}{c}{SM+MSSM} \\
&\multicolumn{1}{c}{$b \rightarrow d$}
&\multicolumn{1}{c}{$\bar{b} \rightarrow \bar{d}$}
&\multicolumn{1}{c}{$b \rightarrow d$}
&\multicolumn{1}{c}{$\bar{b} \rightarrow \bar{d}$} \\
\tableline
$C_{1}^{\text{eff}}$&1.1679+0.0000i&1.1679+0.0000i
&1.1679+0.0000i&1.1679+0.0000i \\
$C_{2}^{\text{eff}}$&$-$0.3525+0.0000i &$-$0.3525+0.0000i
&$-$0.3525+0.0000i&$-$0.3525+0.0000i \\
$C_{3}^{\text{eff}}$&0.0233+0.0018i&0.0250+0.0047i
&0.0188+0.0019i&0.0204+0.0046i\\
$C_{4}^{\text{eff}}$&$-$0.0540$-$0.0053i&$-$0.0588$-$0.0142i
&$-$0.0403$-$0.0057i&$-$0.0451$-$0.0137i \\
$C_{5}^{\text{eff}}$&0.0172+0.0018i&0.0189+0.0047i
&0.0127+0.0019i&0.0143+0.0046i \\
$C_{6}^{\text{eff}}$&$-$0.0675$-$0.0053i&$-$0.0723$-$0.0142i
&$-$0.0539$-$0.0057i&$-$0.0587$-$0.0137i \\
$\tilde{C}_{3}^{\text{eff}}$&&&$-$0.0010+0.0000i&$-$0.0010+0.0000i\\
$\tilde{C}_{4}^{\text{eff}}$&&&0.0029+0.0000i&0.0029+0.0000i\\
$\tilde{C}_{5}^{\text{eff}}$&&&$-$0.0010+0.0000i&$-$0.0010+0.0000i\\
$\tilde{C}_{6}^{\text{eff}}$&&&0.0029+0.0000i&0.0029+0.0000i\\
\end{tabular}
\end{table}

\mediumtext
\begin{table}
\caption{Effective Wilson coefficients $C_{k}^{\text{eff}}(\mu ,q^{2})$ 
for the SM and the combined 
SM+MSSM and effective Wilson coefficients $\tilde{C}_{k}^{\text{eff}}$ 
for the MSSM at 
the renormalization scale $\mu =2.50$ GeV and momentum $q^{2}=m_{b}^{2}/2$
for $b \rightarrow s$ and $\bar{b}
\rightarrow \bar{s}$ transitions. 
Results are given for the MSSM data set (\ref{bqcd37a}).
\label{table4}}
\begin{tabular}{ldddd}
&\multicolumn{2}{c}{SM}&\multicolumn{2}{c}{SM+MSSM} \\
&\multicolumn{1}{c}{$b \rightarrow s$}
&\multicolumn{1}{c}{$\bar{b} \rightarrow \bar{s}$}
&\multicolumn{1}{c}{$b \rightarrow s$}
&\multicolumn{1}{c}{$\bar{b} \rightarrow \bar{s}$} \\
\tableline
$C_{1}^{\text{eff}}$&1.1679+0.0000i&1.1679+0.0000i
&1.1679+0.0000i&1.1679+0.0000i \\
$C_{2}^{\text{eff}}$&$-$0.3525+0.0000i &$-$0.3525+0.0000i
&$-$0.3525+0.0000i&$-$0.3525+0.0000i \\
$C_{3}^{\text{eff}}$&0.0249+0.0032i&0.0248+0.0030i
&0.0202+0.0032i&0.0201+0.0030i\\
$C_{4}^{\text{eff}}$&$-$0.0585$-$0.0095i&$-$0.0583$-$0.0090i
&$-$0.0445$-$0.0095i&$-$0.0442$-$0.0090i \\
$C_{5}^{\text{eff}}$&0.0188+0.0032i&0.0187+0.0030i
&0.0141+0.0032i&0.0140+0.0030i \\
$C_{6}^{\text{eff}}$&$-$0.0721$-$0.0095i&$-$0.0718$-$0.0090i
&$-$0.0581$-$0.0095i&$-$0.0578$-$0.0090i \\
$\tilde{C}_{3}^{\text{eff}}$&&&$-$0.0010+0.0000i&$-$0.0010+0.0000i\\
$\tilde{C}_{4}^{\text{eff}}$&&&0.0031+0.0000i&0.0031+0.0000i\\
$\tilde{C}_{5}^{\text{eff}}$&&&$-$0.0010+0.0000i&$-$0.0010+0.0000i\\
$\tilde{C}_{6}^{\text{eff}}$&&&0.0031+0.0000i&0.0031+0.0000i\\
\end{tabular}
\end{table}

\widetext
\begin{table}
\caption{Comparison of QCD corrected integrated decay rates 
$\Gamma(\mu )$ at scale $\mu $ calculated from (\ref{bqcd49}) and
(\ref{bqcd49a}) for pure penguin
$b \rightarrow q$ and $\bar{b}\rightarrow \bar{q},\;q \in \{d,s\}$ 
transitions 
with the scaled decay rates
$\Gamma^{(0)}(\mu )=\eta^{2}\Gamma^{(0)}(M_{W})$ where $\Gamma^{(0)}(M_{W})$
are the weak scale decay rates calculated using $q^{2}$-dependent 
form factors and  $\eta^{2} \equiv (\alpha_{s}(\mu )/\alpha_{s}(M_{W}))^{2}$ 
accounts for the renormalized $\alpha_{s}$. 
Results are calculated for the cutoff $q^{2}_{\text{min}}=1.0$ GeV$^{2}$  
and for the MSSM data set (\ref{bqcd37a}).
The numbers in parentheses denote powers of 10.
\label{table5}}
\begin{tabular}{lcccccccc}
&\multicolumn{4}{c}{SM}&\multicolumn{4}{c}{SM+MSSM} \\
$\Gamma $(GeV) &$b \rightarrow d$&$\bar{b} \rightarrow \bar{d}$
&$b \rightarrow s$&$\bar{b} \rightarrow \bar{s}$ &
$b \rightarrow d$&$\bar{b} \rightarrow \bar{d}$
&$b \rightarrow s$&$\bar{b} \rightarrow \bar{s}$ \\
\tableline
$\Gamma^{(0)}(M_{W})$& 1.480($-$17)&2.210($-$17)&2.991($-$16)&2.921($-$16) 
&2.062($-$17)&2.926($-$17)&4.080($-$16)&3.997($-$16)\\
&&&&&&&& \\
$\Gamma^{(0)}(m_{b})$&4.898($-$17)&7.311($-$17)&9.897($-$16)&9.667($-$16) 
&6.823($-$17)&9.681($-$17)&1.350($-$15)&1.323($-$15)\\
$\Gamma(m_{b})$& 6.304($-$17)&9.076($-$17)&1.303($-$15)&1.276($-$15) 
&2.931($-$15)&2.825($-$15)&5.054($-$14)&5.065($-$14)\\
&&&&&&&& \\
$\Gamma^{(0)}(2.50)$&7.594($-$17)&1.134($-$16)&1.534($-$15)&1.499($-$15) 
&1.058($-$16)&1.501($-$16)&2.093($-$15)&2.051($-$15)\\
$\Gamma(2.50)$& 8.103($-$17)&1.222($-$16)&1.714($-$15)&1.674($-$15) 
&2.665($-$15)&2.542($-$15)&4.558($-$14)&4.570($-$14)\\
\end{tabular}
\end{table}

\narrowtext
\begin{table}
\caption{SM phase space integrals $I_{i},i=1,2,3$, decay rates  
and branching ratios $B$ for various $b$ decays calculated using
effective Wilson coefficients $C^{\text{eff}}_{k}(\mu ,q^{2})$ evaluated 
at the renormalization scale $\mu $ = 2.50 GeV. The hadronic decay 
rates $\Gamma $ are calculated from (\ref{bqcd50}) using $q^{2}$ fixed  
at $m_{b}^{2}/2$. The numbers in parentheses denote powers of 10.
\label{table6}}
\begin{tabular}{lccccc}
Decay process & $I_{1}$ & $I_{2}$ & $I_{3}$ & $\Gamma $ (GeV) & B  \\
\tableline
$b \rightarrow c\;d\;\bar{u}$ &0.5011&&& 1.846($-$13) &5.007($-$1)\\
$b \rightarrow c\;s\;\bar{u}$ &0.4911&&& 9.280($-$15) &2.517($-$2)\\
$b \rightarrow c\;d\;\bar{c}$ &0.1765&0.1766&8.69($-$2)& 2.921($-$15)
& 7.706($-$3) \\
$b \rightarrow c\;s\;\bar{c}$ &0.1702&0.1706&8.49($-$2) & 5.507($-$14)
& 1.494($-$1) \\
$b \rightarrow u\;d\;\bar{u}$ &1.0000&0.9952&8.38($-$6)& 2.354($-$15)
& 6.386($-$3) \\
$\bar{b} \rightarrow \bar{u}\;\bar{d}\;u$ &&&& 2.586($-$15) & 7.014($-$3)\\
$b \rightarrow u\;s\;\bar{u}$ &0.9884&0.9823&8.34($-$6)& 2.462($-$15) 
& 6.677($-$3) \\
$\bar{b} \rightarrow \bar{u}\;\bar{s}\;u$ &&&& 2.228($-$15) & 6.043($-$3)\\
$b \rightarrow u\;d\;\bar{c}$ &0.4982&&& 6.362($-$17) & 1.726($-$4)\\
$\bar{b} \rightarrow \bar{u}\;\bar{d}\;c$ &&&& 6.362($-$17) & 1.726($-$4)\\
$b \rightarrow u\;s\;\bar{c}$ &0.4882&&& 1.235($-$15) & 3.351($-$3) \\
&&&& \\
$b \rightarrow c\;e^{-}\;\bar{\nu}_{e}$ &0.5012&&& 4.991($-$14) &1.354($-$1) \\
$b \rightarrow c\;\mu ^{-}\;\bar{\nu}_{\mu }$ &0.4982&&& 4.961($-$14) 
&1.346($-$1)\\
$b \rightarrow c\;\tau ^{-}\;\bar{\nu}_{\tau }$ &0.1121&&& 1.117($-$14) 
&3.030($-$2)\\
\end{tabular}
\end{table}

\widetext
\begin{table}
\caption{Experimental and spectator model branching ratios for $B$ meson  
decays. The notation, for example, $B^{0}\rightarrow (D^{-},D^{*\;-})+
(\pi^{+},\rho^{+})$ indicates the branching ratio $B^{0}\rightarrow 
[D^{-}(1.869)\; \text{or} \;D^{*\;-}(2.010)]+(\pi^{+}\; \text{or}\;\rho^{+})$.
The process classification {\em A} or {\em B} refers to whether the 
spectator antiquark combines with the quark $q^{\prime}$, 
Eqn. (\ref{bqcd56}) or
$q$, Eqn. (\ref{bqcd57}). The model classification I or II refers
to no color suppression and total color suppression respectively in 
the formation
of the decay mesons (see the text).
For the decays into light quark systems, the data are for the most part
upper bounds which are not shown. The model predictions are in accord 
with these bounds. Only $\bar{b}$ quark decays are shown, as in the 
PDG tables.
\label{table7}}
\begin{tabular}{lr@{}lcr@{}lr@{}l}
Decay  & \multicolumn{2}{c}{Experiment} & Process & 
\multicolumn{2}{c}{Model I} & 
\multicolumn{2}{c}{Model II} \\
\tableline
$\bar{b} \rightarrow \bar{c}\;l^{+}\;\nu_{l}$ &&&&&&&\\
$B^{+} \rightarrow (\bar{D}^{0},\bar{D}^{*\;0})+l^{+}\nu _{l}$ &
$(7.16 \pm 1.3)$&$\times 10^{-2}$& & $8.6$&$\times 10^{-2}$ &&\\
$B^{0} \rightarrow (D^{-}, D^{*\;-})+l^{+}\nu_{l}$ &
$(6.6 \pm 0.5)$&$\times 10^{-2}$ & & $8.6$&$ \times 10^{-2}$ && \\
$\bar{b} \rightarrow \bar{u}\;l^{+}\;\nu_{l}$&&&&&&& \\
$B^{0} \rightarrow (\pi^{-},\rho^{-})+l^{+}\nu_{l}$ &
$(4.3 \pm 1.6)$&$\times 10^{-4}$& & $5.2$&$ \times 10^{-4}$&& \\
$\bar{b} \rightarrow \bar{c}\;\bar{u}\;d $ &&&&&&& \\
$B^{0} \rightarrow (D^{-},D^{*\;-})+(\pi^{+},\rho^{+})$ &
$(2.04 \pm 0.53)$&$ \times 10^{-2} $& $A$ & $3.00 $&$\times 10^{-2}$ &
$ 2.73 $&$\times 10^{-2}$ \\
$B^{+} \rightarrow (\bar{D}^{0},\bar{D}^{*\;0})+(\pi ^{+},\rho ^{+})$ &
$(3.88 \pm 0.58 )$&$\times 10^{-2}$ & $A$ & $3.00 $&$\times 10^{-2}$ &
$2.78 $&$\times 10^{-2}$ \\
&& & $B$ & $1.39$&$ \times 10^{-2}$ & $1.55$&$ \times 10^{-5}$ \\
$B^{0} \rightarrow (\bar{D}^{0},\bar{D}^{*\;0})+(\pi ^{0},\eta , \rho^{0},
\omega ) $ & $< 3.2 $&$\times 10^{-3} $ & $B$ & $1.39$&$ \times 10^{-2} $ &
$1.55 $&$\times 10^{-5} $ \\
$\bar{b} \rightarrow \bar{c}\;\bar{s}\;c$ &&&&&&& \\
$B^{+} \rightarrow (\bar{D}^{0},\bar{D}^{*\;0})+(D^{+}_{s},D^{*\;+}_{s})$ &
$(6.1 \pm 2.3)$&$ \times 10^{-2} $ & $A$ & $3.2$&$ \times 10^{-2}$ & 
$2.9$&$ \times 10^{-2} $ \\
$B^{0} \rightarrow (D^{-},D^{*\;-})+(D^{+}_{s},D^{*\;+}_{s})$ &
$(4.8 \pm 1.8 )$&$ \times 10^{-2}$ & $A$ & $3.2 $&$\times 10^{-2} $ &
$2.9 $&$\times 10^{-2} $ \\
$B^{0} \rightarrow (\psi (1s),\chi (1P))+(K^{0},K^{*\;0}) $ &
$ < 6$&$ \times 10^{-3}$ & $B$ & $3.7$&$ \times 10^{-2}$ & 
$4.1 $&$\times 10^{-5}$ \\
$B^{+} \rightarrow (\psi (1s),\chi (1P))+(K^{+},K^{*\;+})$ && & $B$ & 
$ 3.7 $&$\times 10^{-2}$ & $ 4.1 $&$\times 10 ^{-5}$ \\
$ \bar{b} \rightarrow \bar{u}\;\bar{d}\;u $ &&&&&&& \\
$B^{0} \rightarrow (\pi ^{+},\rho ^{+})+(\pi ^{-},\rho ^{-})$ &
&& $A$ & $1.07 $&$\times 10^{-4}$ & $0.95 $&$\times 10^{-4} $ \\
$B^{+} \rightarrow (\pi ^{+},\rho^{+})+(\pi ^{0},\eta ,\rho^{0}, \omega ) $ &
&& $ A $ & $1.07$&$ \times 10^{-4}$ & $0.95 $&$\times 10^{-4}$ \\
&& & $B$ & $1.09 $&$\times 10^{-4}$ & $ 1.9 $&$\times 10^{-7} $ \\
$B^{0} \rightarrow (\pi ^{0},\eta , \rho^{0},\omega )+(\pi ^{0}, \eta ,
\rho^{0}, \omega ) $ && & $B$ & $ 1.09 $&$\times 10^{-4}$ & 
$ 1.9$&$ \times 10^{-7} $ \\
$b \rightarrow u\;d\;\bar{u} $ &&&&&&& \\
$\bar{B}^{0} \rightarrow (\pi^{+},\rho^{+})+(\pi ^{-},\rho^{-})$ & &
&$A$ & $0.95$&$ \times 10^{-4}$ & $ 0.87 $&$\times 10^{-4} $ \\
$B^{-} \rightarrow (\pi ^{-},\rho^{-})+(\pi ^{0},\eta , \rho^{0},\omega ) $ &
&& $A$ & $ 0.95$&$ \times 10^{-4} $ & $ 0.87$&$ \times 10^{-4}$ \\
&&& $B$ & $ 0.97 $&$\times 10^{-4}$ & $1.9$&$ \times 10^{-7}$ \\
$ \bar{B}^{0}\rightarrow (\pi ^{0},\eta , \rho^{0},\omega ) +(\pi ^{0},
\eta , \rho^{0}, \omega )$ && & $B$ & $ 0.97 $&$\times 10^{-4}$ & 
$1.9 $&$\times 10^{-7}$ \\
$\bar{b} \rightarrow \bar{u}\;\bar{s}\;u $ &&&&&&& \\
$B^{0} \rightarrow (K^{+},K^{*\;+})+(\pi ^{-}, \rho^{-})$ &
&& $A$ &$ 0.503 $&$\times 10^{-4} $ & $ 0.422$&$ \times 10^{-4} $ \\
$B^{+} \rightarrow (K^{+},K^{*\;+})+(\pi ^{0},\eta , \rho^{0}, \omega )$ &
&& $A$ & $ 0.503$&$ \times 10^{-4}$ & $ 0.422 $&$\times 10^{-4} $ \\
&& & $B$ & $1.09 $&$\times 10^{-4}$ & $0.93 $&$\times 10^{-6}$ \\
$B^{0} \rightarrow (K^{0},K^{*\;0})+(\pi ^{0},\eta , \rho^{0}, \omega ) $ &
&& $B$ & $1.09 $&$\times 10^{-4}$ & $0.93$&$ \times 10^{-6} $ \\
$b \rightarrow u\;s\;\bar{u}$ &&&&&&& \\
$\bar{B}^{0} \rightarrow (K^{-},K^{*\;-})+(\pi ^{+},\rho^{+}) $ && & $A$ &
$0.624$&$ \times 10^{-4}$ & $0.531$&$ \times 10^{-4} $ \\
$B^{-} \rightarrow (K^{-},K^{*\;-})+(\pi^{0},\eta , \rho^{0}, \omega )$ &
&& $A$ & $0.624 $&$\times 10^{-4} $ & $0.531 $&$\times 10^{-4}$ \\
&& & $B$ & $ 1.21 $&$\times 10^{-4}$ & $0.93 $&$\times 10^{-6} $ \\
$ \bar{B}^{0} \rightarrow (\bar{K}^{0},\bar{K}^{*\;0}) +(\pi ^{0}, \eta ,
\rho^{0}, \omega ) $ & && $B$ & $1.21 $&$\times 10^{-4} $ & 
$ 0.93$&$ \times 10^{-6}$ \\
$\bar{b} \rightarrow \bar{d}\;\bar{s}\;d$ &&&&&&& \\
$B^{+} \rightarrow (K^{0},K^{*\;0})+(\pi^{+},\rho^{+}) $ && & $A$ &
$0.491 $&$\times 10^{-4}$ & $0.410 $&$\times 10^{-4}$ \\
$B^{0} \rightarrow (K^{0},K^{*\;0})+(\pi^{0},\eta , \rho^{0}, \omega )$ &&
& $A$ & $0.491$&$ \times 10^{-4}$ & $0.410 $&$\times 10^{-4}$ \\
&& & $B$ & $1.082$&$ \times 10^{-4}$ & $0.920 $&$\times 10^{-6} $ \\
$b \rightarrow d\;s\;\bar{d}$ &&&&&& &\\
$B^{-} \rightarrow (\bar{K}^{0},\bar{K}^{*\;0})+(\pi^{-},\rho^{-}) $ && & 
$A$ &$0.499 $&$\times 10^{-4}$ & $0.417$&$ \times 10^{-4}$ \\
$\bar{B}^{0} \rightarrow (\bar{K}^{0},\bar{K}^{*\;0})+(\pi^{0},\eta , \rho^{0}, \omega )$ &
& &$A$ & $0.499 $&$\times 10^{-4}$ & $0.417$&$ \times 10^{-4}$ \\
& && $B$ & $1.096 $&$\times 10^{-4}$ & $0.914 $&$\times 10^{-6} $ \\

\end{tabular}
\end{table}

\end{document}